\documentclass[pra,aps,reprint,amsmath,amssymb,showpacs,floatfix,superscriptaddress]{revtex4-1}

\usepackage[english]{babel}
\usepackage[utf8]{inputenc}
\usepackage[colorinlistoftodos]{todonotes}
\usepackage[colorlinks,citecolor=blue,linkcolor=blue]{hyperref}
\usepackage{graphicx}	
\usepackage{dcolumn}	
\usepackage{bm}			
\usepackage{txfonts} 	

\newcommand{\CNN}{Centre de Nanosciences et de Nanotechnologies, CNRS, Univ. Paris-Sud, Universit{\'e} Paris-Saclay, 91405 Orsay, France}
\newcommand{\UMPhy}{Unit{\'e} Mixte de Physique, CNRS, Thales, Univ. Paris-Sud, Universit{\'e} Paris-Saclay, 91767 Palaiseau, France}

\begin{document}

\title{Current-driven skyrmion expulsion from magnetic nanostrips}

\author{Myoung-Woo Yoo}
\email{myoung-woo.yoo@u-psud.fr}
\affiliation{\CNN}
\affiliation{\UMPhy}
\author{Vincent Cros}
\affiliation{\UMPhy}
\author{Joo-Von Kim}
\email{joo-von.kim@u-psud.fr}
\affiliation{\CNN}

\date{\today}

\begin{abstract}
We study the current-driven skyrmion expulsion from magnetic nanostrips using micromagnetic simulations and analytic calculations. We explore the threshold current density for the skyrmion expulsion, and show that this threshold is determined by the critical boundary force as well as the spin-torque parameters. We also find the dependence of the critical boundary force on the magnetic parameters; the critical boundary force decreases with increasing the exchange stiffness and perpendicular anisotropy constants, while it increases with increasing Dzyaloshinskii-Moriya interaction and saturation magnetization constants. Using a simple model describing the skyrmion and locally-tilted edge magnetization, we reveal the underlying physics of the dependence of the critical boundary force on the magnetic parameters based on the relation between the scaled Dzyaloshinskii-Moriya-interaction parameter and the critical boundary force. This work provides a fundamental understanding of the skyrmion expulsion and the interaction between the skymion and boundaries of devices and shows that the stability of the skyrmion in devices can be related to the scaled Dzyaloshinskii-Moriya-interaction parameter of magnetic materials.
\end{abstract}

\maketitle

\section{Introduction}
Magnetic skyrmions are non-trivial magnetic configurations that are stabilized by presence of the Dzyaloshinskii-Moriya interaction (DMI) \cite{Dzyaloshinsky:1958br, Moriya:1960go, Moriya:1960kc, Fert:1980hr, Fert:1990bn, Crepieux:1998fj, Fert:2013fq}. Skyrmions have vortex- or hedgehog-like two-dimensional configurations at the nanometer scale in perpendicular magnetization systems and are stable under specific conditions due to their topology. They have been predicted to occur in non-centrosymmetric crystals or ultrathin films lacking inversion symmetry~\cite{Bogdanov:1989vd, Bogdanov:1999df}, and have been recently observed in chiral-lattice magnets and heavy-metal/ultrathin-ferromagnet heterostructures at room temperature \cite{Muhlbauer:2009bc, Pappas:2009bk, Yu:2010iu, Heinze:2011ic, Tokunaga:2015cf, MoreauLuchaire:2016em, Jiang:2015csa, Woo:2016jw, Boulle:2016jt, Hrabec:2017wy}. Skyrmions have been studied intensively over the past few years because they exhibit interesting features, such as the topological Hall effect, one aspect of the emergent electrodynamics~\cite{Everschor:2011cc, Schulz:2012bi, Nagaosa:2013cc, EverschorSitte:2014dq}. Very recently, a topology-induced Hall-like behavior of isolated skyrmions, so-called skyrmion Hall effect, have been observed by magneto-optical Kerr microscopy and time-resolved X-ray microscopy.~\cite{Jiang:2016ha, Litzius:2016jp}.

Skyrmions have also attracted much attention because of their potential applications for more efficient data storage, as information carriers, and  for microwave oscillators~\cite{Kiselev:2011cm, Fert:2013fq, Sampaio:2013kn, Zhang:2015hx, GarciaSanchez:2016cxa}. Most of these applications rely on current-driven motion in confined geometries. The potential performance of such devices is related to how quickly and reliably a skyrmion can be propagated within the nanostructure, which ultimately depends on the current densities applied~\cite{Everschor:2011cc, Sampaio:2013kn, Nagaosa:2013cc, Iwasaki:2013ji, EverschorSitte:2014dq, Tomasello:2014kaa, Jiang:2016ha}. However, there exists a threshold current density above which the skyrmion can be expelled from the nanostructure at the boundary edges, which places a severe constraint on the upper limit for skyrmion propagation speeds that can be attained using currents~\cite{Sampaio:2013kn, Iwasaki:2014hb, Iwasaki:2014dk}, if the system is not specially designed to prevent the skyrmion reaching the boundaries~\cite{Barker:2016fq, Zhang:2016kf,Muller:2017hb}. It is therefore desirable to have a quantitative understanding of the conditions under which such expulsion occurs, although there were several earlier studies focused on the interaction between current-driven skyrmions and boundaries of magnetic confinements~\cite{Sampaio:2013kn,Iwasaki:2014hb, Iwasaki:2014dk,Zhang:2015kra}.

In this article, we present a theoretical investigation of the expulsion of a skyrmion when moving through spin torque in a nanostructure. First, using micromagnetic simulations, we evaluate the threshold current densities for the expulsion with different spin-torque parameters, the Gilbert damping and/or non-adiabaticity parameters. Based on the simulation results, we calculate a critical boundary force that is a key parameter for determining the critical current density, and obtain the dependence of the critical force on the magnetic parameters, such as exchange stiffness, perpendicular anisotropy, saturation magnetization, and DMI constants. Finally, using an analytical model, we examine the underlying physics of the relation between the critical boundary force and the magnetic parameters.

\section{Geometry and simulation method}
\begin{figure}
\centering
\includegraphics[width=8.6cm]{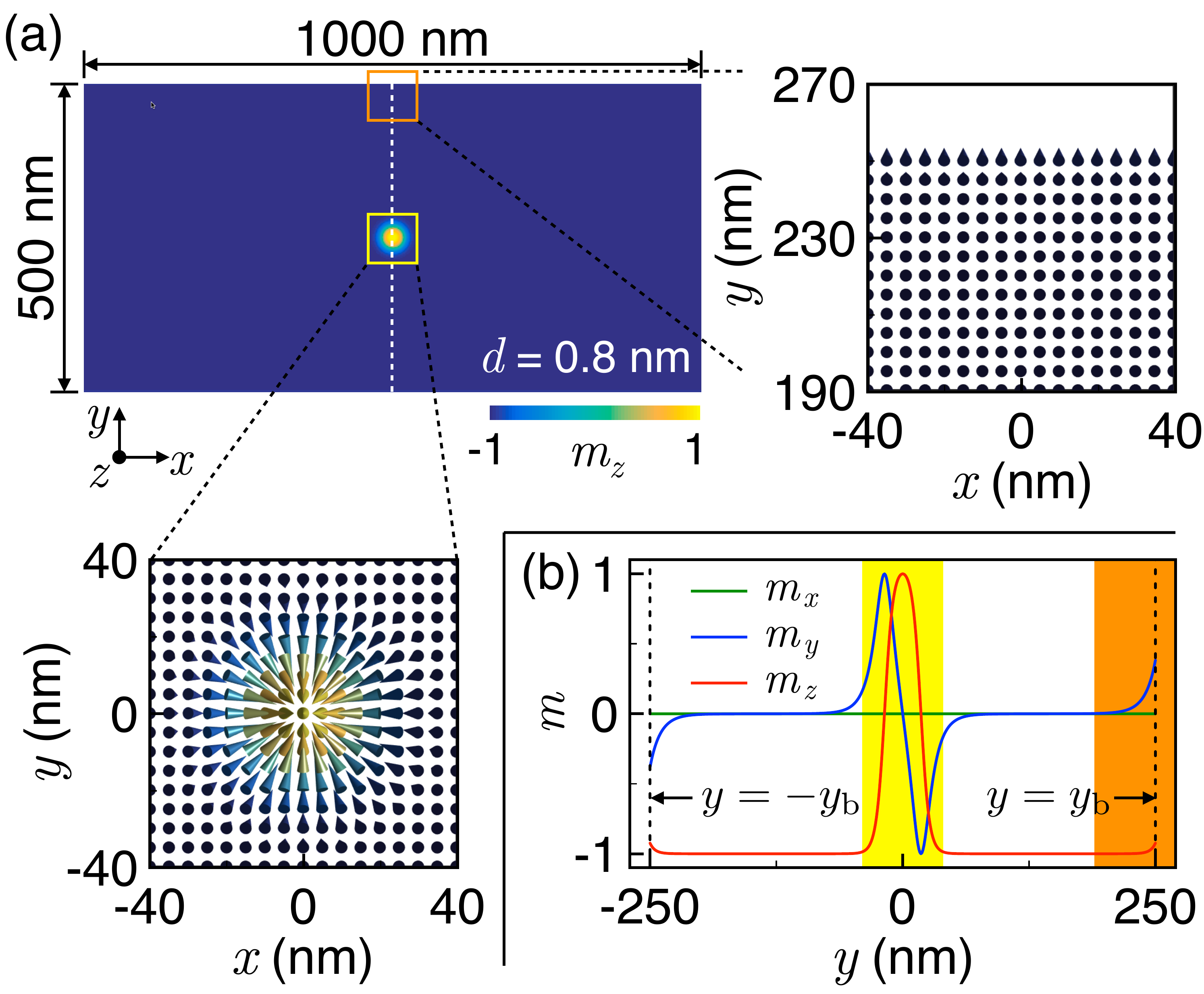}
\caption{\label{Fig:InitialState}
(a) A nanostrip model for micromagnetic simulations, where $d$ is the thickness of the film. The color indicates an initial $m_{z}$ configuration in which a meta-stable skyrmion is located at the center of the wire.  The two insets show local magnetization vectors of the skyrmion (yellow-colored region) and an edge of the nanostrip (orange-colored region), respectively. Note that the moments are drawn only for about 41$\%$ of the effectively simulated sites. (b) $m_x$, $m_y$, and $m_z$-profiles crossing the skyrmion center parallel to the $y$-axis (the white dashed-line in (a)). Black dashed lines indicate boundaries of the nanowire, $y$ = $-y_\mathrm{b}$ and $y_{\mathrm{b}}$, where, $y_\mathrm{b}$ = 250 nm. The yellow- and orange-colored regions correspond to the yellow- and orange-colored regions in (a), respectively.
}
\end{figure}
The MuMax3 code is used for micromagnetic simulations~\cite{Vansteenkiste:2014et}. A 1000 $\times$ 500 $\times$ $d$ $\mathrm{nm^3}$ nanowire is chosen with $d$ = 0.8 nm and the system is uniformly discretized with 512 $\times$ 256 $\times$ 1 finite difference cells [Fig.~\ref{Fig:InitialState} (a)]. Periodic boundary conditions are used for the $x$-direction to mimic an infinitely long nanostrip. We consider the dipolar interaction in the simulations. The magnetic parameters used here correspond to those of a Pt/Co/Ir multilayer~\cite{MoreauLuchaire:2016em}; we consider an exchange stiffness constant of $A_\mathrm{ex}$ = 16 $\times$ $10^{-12}$ $\mathrm{J/m}$, a perpendicular anisotropy constant of $K_\mathrm{u}$ = 0.717 $\times$ $10^{6}$ $\mathrm{J/m^3}$, and a saturation magnetization of $M_\mathrm{s}$ = 0.956 $\times$ $10^6$ A/m. An interfacial DMI constant of $D_\mathrm{i}$ = 1.5 $\mathrm{mJ/m^{2}}$ is chosen such that the isolated skyrmion state is metastable. Figure \ref{Fig:InitialState} shows the initial state of the micromagnetic simulations obtained by the energy minimization method, in which the N{\'e}el-type skyrmion at the center and DMI-induced locally-tilted magnetization near the edges are present~\cite{Sampaio:2013kn, Rohart:2013ef, GarciaSanchez:2014dw}.

For the current-driven dynamics, we solve the Landau-Lifshitz equation with Gilbert damping and consider separately spin-torques, $\Gamma_{\rm st}$, associated with current flowing in the film plane (CIP) or perpendicular to the film plane (CPP)~\cite{Gilbert:2004aa, Zhang:2004hs, Slonczewski:1996gq},
\begin{equation}
\frac{d \mathbf{m}}{dt} = -\gamma_0 \mathbf{m} \times \mathbf{H}_{\rm eff} + \alpha \mathbf{m} \times \frac{d \mathbf{m}}{dt} + \Gamma_{\rm st}.
\end{equation}
This equation describes the time evolution of the magnetization configuration described the unit vector $\mathbf{m} = \mathbf{m}(\mathbf{r},t)$, where $\gamma_0 = \mu_0 \gamma$ is the gyrotropic ratio, $\mathbf{H}_{\rm eff}$ is the effective field and $\alpha$ is the Gilbert damping constant. For the CIP case, we use the Zhang-Li form for the spin torques~\cite{Zhang:2004hs, Vansteenkiste:2014et}, 
\begin{equation}
\Gamma_{\rm st, CIP} = -\left( \mathbf{v}_{\rm s} \cdot \nabla \right)\mathbf{m} + \beta \mathbf{m} \times \left( \mathbf{v}_{\rm s} \cdot \nabla \right)\mathbf{m},
\end{equation}
where $\mathbf{v}_{\rm s}$ is an effective spin-current drift-velocity with a magnitude of $v_{\rm s} = -\mu_\mathrm{B} p/\left[ e M_s ( 1 + \beta^2) \right] j$, $\mu_\mathrm{B}$ is the Bohr magneton, $j$ is the current density, and $\beta$ is the nonadiabatic spin torque parameter. We assume a spin polarization of $p = 0.5$ for all simulations. For the CPP case, we use the Slonczewski form for the spin torques~\cite{Slonczewski:1996gq, Vansteenkiste:2014et}, 
\begin{equation}
\Gamma_{\rm st, CPP} = \zeta j \mathbf{m} \times (\mathbf{m} \times \hat{\mathbf{p}}),
\end{equation}
where $\zeta = \gamma \hbar p/(2 e M_\mathrm{s} d)$ is the efficiency factor and $\hat{\mathbf{p}}$ is the unit vector of the spin polarization. This term is equivalent to the spin torque induced by the spin Hall effect, if we set $j = j_\mathrm{hm}$ and $p = \vartheta_\mathrm{sh}$, where $j_\mathrm{hm}$ and $\vartheta_\mathrm{sh}$ are the current density flowing in the heavy metal and the spin Hall angle, respectively~\cite{Taniguchi:2015ch}.

\section{Results}

\subsection{Micromagnetics simulations}
\begin{figure}
\centering\includegraphics[width=8.6cm]{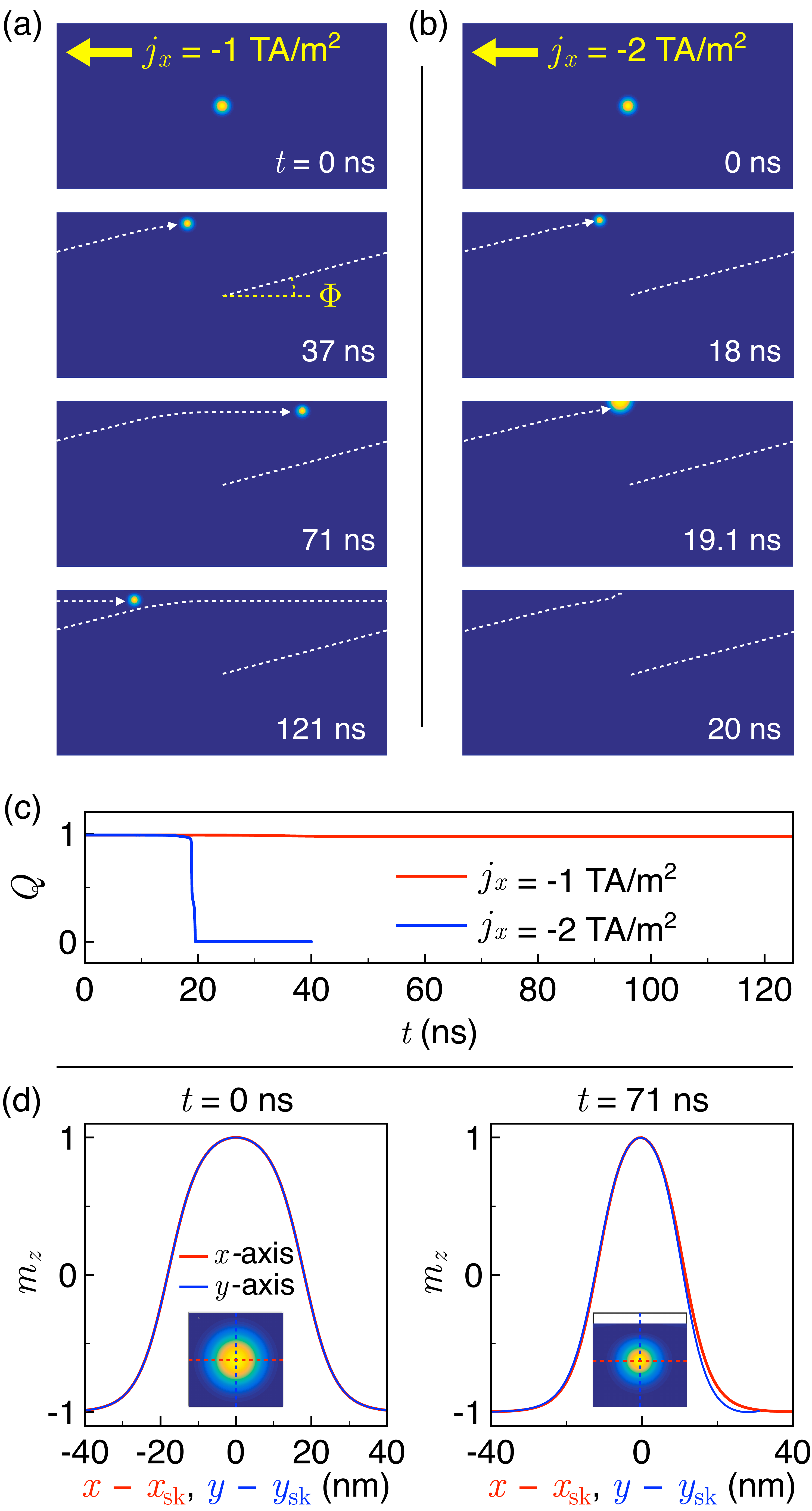}
\caption{\label{Fig:Trajectory}
(a) Snapshots of $m_z$ configuration during the $j_x$-driven skyrmion-motion at indicated time, $t$, for $\alpha$ = 0.3 and $\beta$ = 0.1. The applied current density, $j_x = -1.0 \times 10^{12}$ $\mathrm{A/m^2}$, is smaller than the threshold current density, $j_x^\mathrm{th}$. (b) Snapshots of $m_z$ configuration during the $j_x$-driven skyrmion-motion up to the the skyrmion expulsion with $j_x = -2.0 \times 10^{12}$ $\mathrm{A/m^2}$ which is larger than $j_x^\mathrm{th}$. White dashed lines in (a) and (b) show trajectories of the skyrmion from $t$ = 0. $\Phi$ is an angle between the $x$-axis and the trajectory at $t$ $\sim$ 0 ns. (c) Evolutions of the topological charge, $Q$, over the system in the cases of $j_x = -1.0 \times 10^{12}$ $\mathrm{A/m^2}$ (red line) and $j_x = -2.0 \times 10^{12}$ $\mathrm{A/m^2}$ (blue line). (d) $m_z$ profiles parallel to $x$- (red lines) and $y$-axes (blue lines) at $t$ = 0 ns and 71 ns during the skyrmion motion with $j_x = -1.0 \times 10^{12}$ $\mathrm{A/m^2}$. $\mathbf{X}_\mathrm{sk} (t) = \left( x_{\mathrm{sk}} (t), y_{\mathrm{sk}} (t) \right)$ is the position vector of the skyrmion center in the nanostrip. The insets show the $m_z$ configurations near the skyrmion at indicated times in (a), and the dashed lines corresponds to the paths of the profiles.
}
\end{figure}
Using micromagnetic simulations, the threshold values of the in-plane current density, $j_x^\mathrm{th}$, are extracted for different values of the Gilbert damping constant, $\alpha$, and the nonadiabatic spin-torque parameter, $\beta$, because these parameters determine the current-driven motion of the skyrmion with respect to the direction of the current flow ~\cite{Everschor:2011cc}. After the application of the $x$-directional in-plane current, $j_x$, the skyrmion starts to move from the initial position, $\mathbf{X}_\mathrm{sk} = \left( x_\mathrm{sk}, y_\mathrm{sk} \right) = \left( 0, 0 \right)$, in a diagonal direction with an angle of $\Phi$ [Fig. \ref{Fig:Trajectory}(a)], because of the skyrmion Hall effect. This motion corresponds to the dynamics in infinite films where boundary edges are not present. After a certain duration, the skyrmion reaches one of the boundaries of the nanostrip, which for applied currents below a threshold $j_x < j_x^\mathrm{th}$, the skyrmion exhibits only motion along the $x$-direction at constant $y_\mathrm{sk}$ as a result of the restoring force induced by the boundary, $\mathbf{F}_\mathrm{b}$ \cite{Sampaio:2013kn,Iwasaki:2014hb, Iwasaki:2014dk}. In this case, the topological charge over the total system, $Q = \left(1/4\pi\right) \int \mathbf{m} \cdot \left( \partial_x \mathbf{m} \times \partial_y \mathbf{m} \right) \mathrm{d}x \mathrm{d}y$, is conserved, as shown in Fig. \ref{Fig:Trajectory}(c) (red line). This motion is accompanied by a reduction in the size of the skyrmion core, but the radial symmetry of the spin configuration about the core center is largely preserved [Fig.~\ref{Fig:Trajectory}(d)]. On the other hand, in the case of $j_x > j_x^\mathrm{th}$, the skyrmion is annihilated after shrinking [Fig. \ref{Fig:Trajectory}(c)], then expelled from the nanostrip ($t$ = 18 ns), as shown in Fig. \ref{Fig:Trajectory}(b).

\begin{figure}
\centering\includegraphics[width=8.6cm]{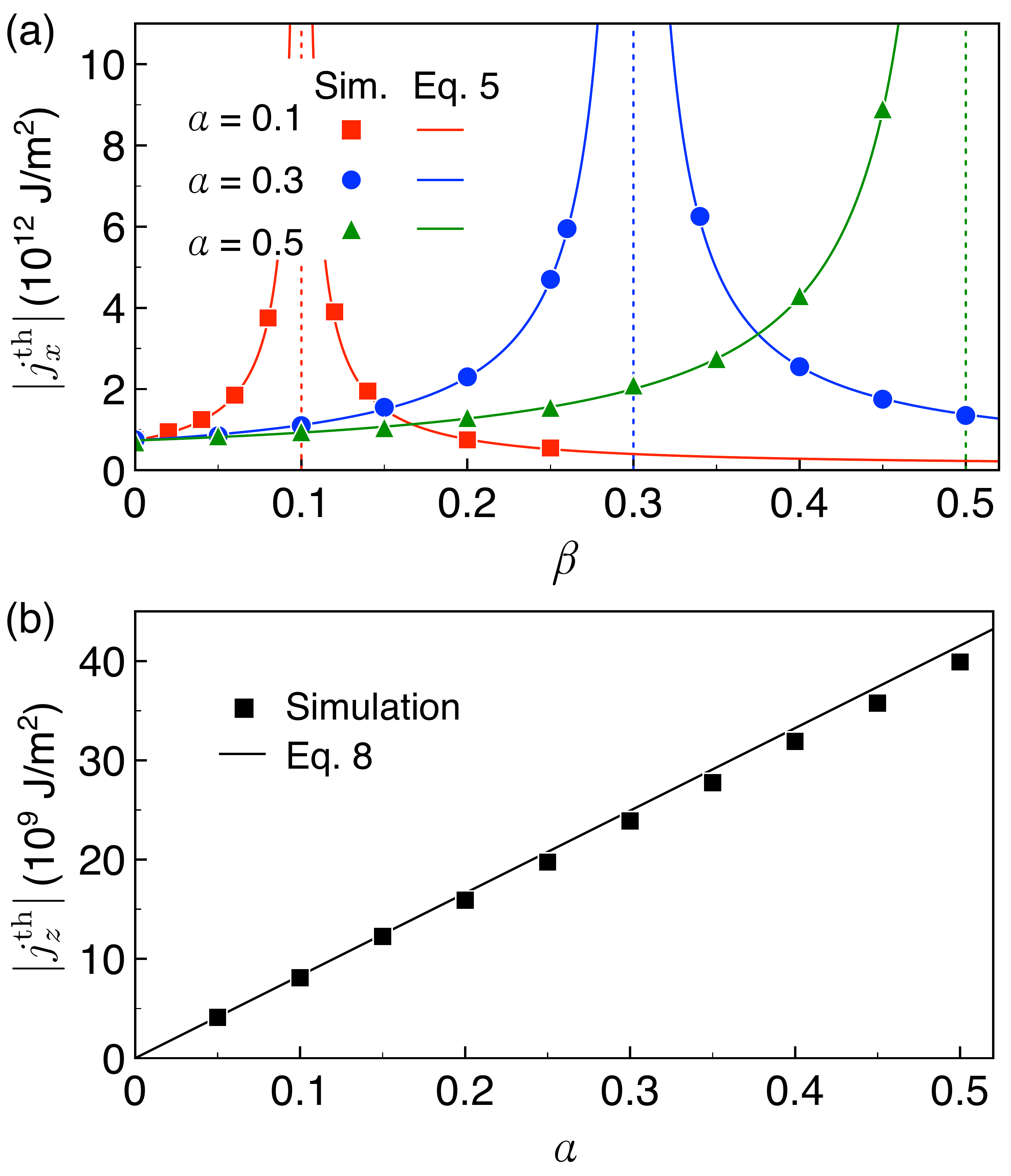}
\caption{\label{Fig:jth}
(a) $\left| j_x^\mathrm{th} \right|$ versus $\beta$ for different values of $\alpha$ obtained from the micromagnetic simulations (symbols) and Eq. \ref{Eq:jxth} (solid lines). The dashed-lines show $\beta$ = $\alpha$. (b) $\left| j_z^{\mathrm{th}} \right|$ versus $\alpha$ obtained from the micromagnetic simulations (symbol) and Eq. \ref{Eq:jzth} (solid line). Error bars of the simulations are the same size or smaller than the symbols. For the analytic calculations in (a) and (b), $F_\mathrm{b}^\mathrm{c}$ = 1.215 $\times$ $10^{-12}$ N is used.
}
\end{figure}

From the simulations, we obtained $j_x^\mathrm{th}$ in a wide range of different $\alpha$ and $\beta$ values~\cite{Schellekens:2013it,Sampaio:2013kn}. In Fig.~\ref{Fig:jth}(a) (symbols), we show the dependence of this threshold current as a function of $\beta$, which is presented for three different values of $\alpha$. 
The threshold current is found to diverge when $\beta$ approaches $\alpha$ and the curves are largely symmetric about $\beta$ = $\alpha$. This divergence at $\beta$ = $\alpha$ (dashed lines) results from the fact that the deflection angle vanishes for this choice of parameters ($\Phi = 0$).

Similar motion and expulsion can also be achieved by spin polarized currents in the CPP geometry~\cite{Sampaio:2013kn}, $j_z$, for which a finite threshold is also found. In this geometry, the deflection angle $\Phi$ depends only on the Gilbert damping constant $\alpha$, in contrast to the CIP geometry for which it is the ratio between $\alpha$ and the nonadiabaticity $\beta$ that counts. The variation of the threshold current as a function of $\alpha$ is presented in Fig.~\ref{Fig:jth}(b) (symbols), where a linear relationship is found. Here, $j_z$ is assumed to be spin polarized in the $+y$-direction, i.e., $\hat{\mathbf{p}} = +\hat{\mathbf{y}}$.

\subsection{Analytical model of the critical boundary force}
\begin{figure}
\centering
\includegraphics[width=8.6cm]{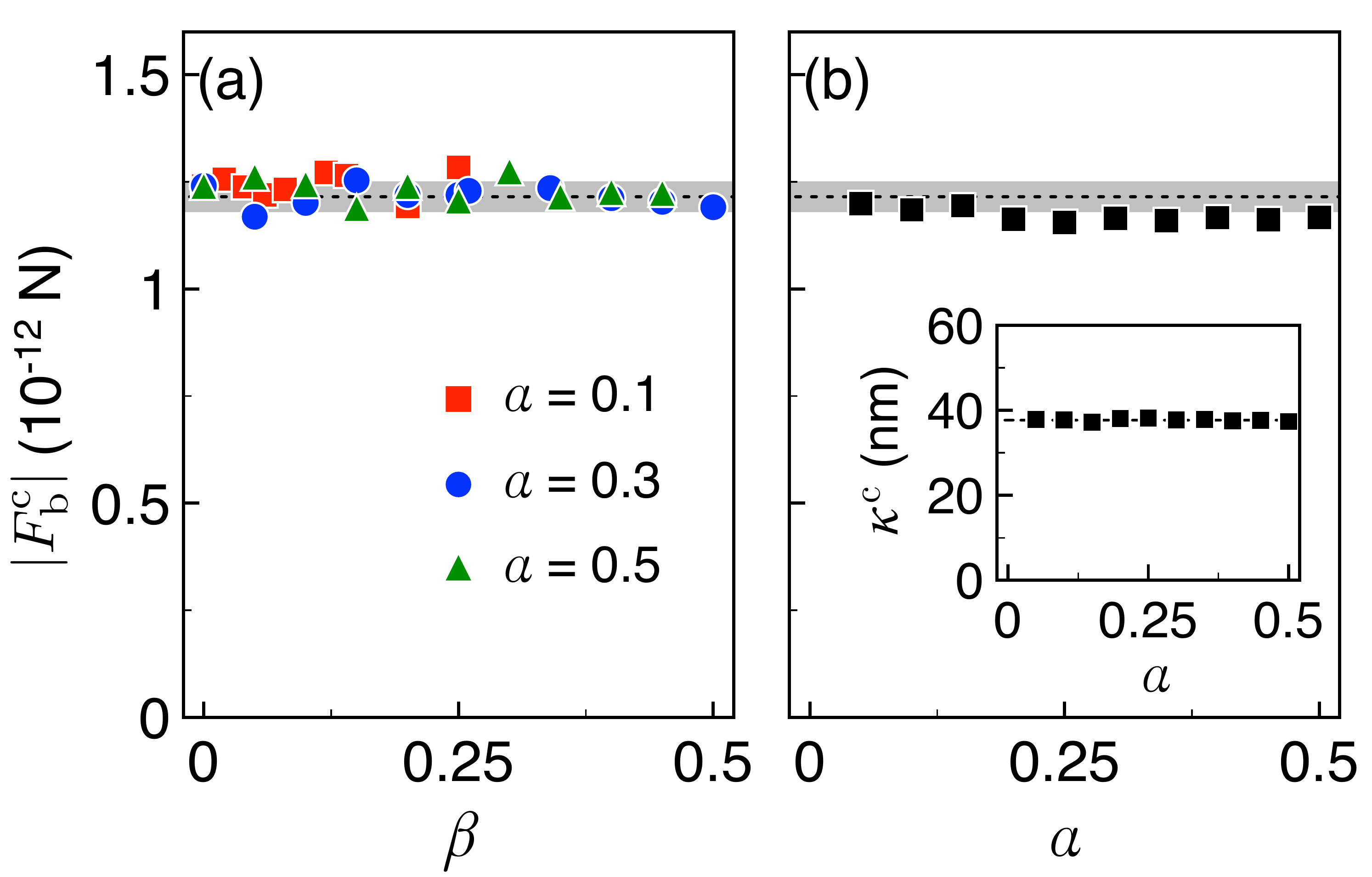}
\caption{\label{Fig:Fbc}
(a) $\left| F_\mathrm{b}^\mathrm{c} \right|$ versus $\beta$ for different values of $\alpha$ in the case of the $j_x$-driven skyrmion expulsion (Eq. \ref{Eq:jxth}). (b) $\left| F_\mathrm{b}^\mathrm{c} \right|$ versus $\alpha$ in the case of the $j_z$-driven skyrmion expulsion (Eq. \ref{Eq:jzth}). The dashed lines and gray-colored regions  in (a) and (b) indicate the mean value and standard deviation, respectively, $\left| F_{\mathrm{b}}^{\mathrm{c}} \right|$ = (1.215 $\pm$ 0.036) $\times$ $10^{-12}$ N. The inset in (b) shows the critical characteristic-size for the expulsion, $\kappa_\mathrm{c}$, and the dashed line indicates the mean value of $\kappa_\mathrm{c}$.
}
\end{figure}

Based on the simulation results, we investigated the underlying physics of the threshold current density by using Thiele's approach~\cite{Thiele:1973br}, which involves assuming a rigid profile for the skyrmion that allows us to integrate out all other degrees of freedom. As such, the approach allows us to describe the dynamics entire in terms of the skyrmion position, $\mathbf{X}_\mathrm{sk}$. In order to analyze the $j_x$-driven steady-state skyrmion-motion near the edge, we assume a Thiele equation of the following form,
\begin{equation}
\mathbf{G} \times (\mathbf{v}_{\rm s} - \mathbf{v}) + \mathcal{D}(\beta \mathbf{v}_{\rm s} - \alpha \mathbf{v}) + \mathbf{F}_{\mathrm{b}} = 0,
\end{equation}
where $\mathbf{F}_\mathrm{b} = \partial U / \partial \mathbf{X}_\mathrm{sk}$ is the boundary force. Here, $U$ is the total magnetic energy of the system, $\mathbf{v} = (v_x, v_y, 0)$ is the skyrmion velocity, $\mathbf{G} = \hat{\mathbf{z}}G = \hat{\mathbf{z}}(4\pi Q)M_sd/\gamma$ is the gyrovector, $Q$ is topological charge of a skyrmion, $\gamma$ is the gyromagnetic ratio, and $\mathcal{D} = -(16/3)\pi M_{\mathrm{s}}d/\gamma $ is a damping constant~\cite{Everschor:2011cc, Iwasaki:2014dk, Kim:2015bv}. In the geometry we consider, $\mathbf{F}_\mathrm{b}$ only has a $y$-component, i.e., $\mathbf{F}_\mathrm{b} = \hat{\mathbf{y}} F_\mathrm{b}$, by assuming infinitely long nanostrips in the $x$-direction. Because of the force balance, $\left| F_\mathrm{b} \right|$ increases with increasing $j_x$, and $\left| F_\mathrm{b} \right|$ reaches the maximum value, $\left| F_\mathrm{b}^\mathrm{c} \right|$ at $j_x = j_x^\mathrm{th}$. The analytic form of $j_x^\mathrm{th}$ can be obtained as
\begin{equation} \label{Eq:jxth}
	\frac{1}{j_x^\mathrm{th}}=-\frac{G\tau}{F_\mathrm{b}^\mathrm{c}} \left(1-\frac{\beta}{\alpha}\right),
\end{equation}
where we have assumed $v_y = 0$ near the edge and inserted $j_x$ = $j_x^\mathrm{th}$ and $F_\mathrm{b}$ = $F_\mathrm{b}^\mathrm{c}$. Here, $\tau = -\mu_\mathrm{B} p/\left[ e M_\mathrm{s} ( 1 + \beta^2) \right]$. Equation~(\ref{Eq:jxth}) clearly shows that $j_x^\mathrm{th}$ is a function of $F_\mathrm{b}^\mathrm{c}$ as well as $\alpha$ and $\beta$. As such, $F_{\mathrm{b}}^\mathrm{c}$ can be calculated by rearranging Eq.~(\ref{Eq:jxth})
\begin{equation} \label{Eq:Fbc}
	F_\mathrm{b}^\mathrm{c} = -G \tau \left( 1- \frac{\beta}{\alpha} \right) j_x^\mathrm{th},
\end{equation}
which can be obtained numerically by using the values of $j_x^\mathrm{th}$ determined from simulations in Fig.~\ref{Fig:jth}(a). The calculated values of $\left| F_\mathrm{b}^\mathrm{c} \right|$ are plotted in Fig. \ref{Fig:Fbc}(a) for the different $\alpha$ and $\beta$ considered, and we find that $\left| F_\mathrm{b}^\mathrm{c} \right|$ does not depend on $\alpha$ and $\beta$, unlike $j_x^\mathrm{th}$.

$F_\mathrm{b}^\mathrm{c}$ for $j_z$-driven skyrmion expulsion was also examined. A similar Thiele equation can be obtained for this geometry, 
\begin{equation}
\mathbf{G} \times \mathbf{v} + \alpha \mathcal{D} \mathbf{v} + \mathbf{F}_\mathrm{st} + \mathbf{F}_\mathrm{b} = 0,
\end{equation}
where $\mathbf{F}_\mathrm{st} = \sigma \kappa j_{z} \hat{\boldsymbol{z}} \times \hat{\mathbf{p}}$  is a force from the spin torque with $\sigma = -\pi \hbar p/(2e)$, and $\kappa = \int_{0}^{\infty} dr \left( r \partial_r \theta + \sin{\theta} \cos{\theta} \right)$ is the characteristic length of the skyrmion \cite{Sampaio:2013kn, GarciaSanchez:2016cxa}. $\theta$ and $r$ are the polar angle of the local magnetization and the distance from the skyrmion center, respectively. By assuming $v_y = 0$ and inserting $j_z = j_z^\mathrm{th}$ and $F_\mathrm{b}$ = $F_\mathrm{b}^\mathrm{c}$, an analytic form of $j_{z}^\mathrm{th}$ can be obtained as a function of $\alpha$ and $F_{\mathrm{b}}^\mathrm{c}$ as
\begin{equation} \label{Eq:jzth}
	j_z^\mathrm{th} = -\frac{\mathcal{D} F_\mathrm{b}^\mathrm{c}}{G\sigma \kappa^\mathrm{c}} \alpha,
\end{equation}
where $\kappa^\mathrm{c}$ is the size of the compressed skyrmion before expulsion. $\kappa^\mathrm{c}$ can be determined numerically from the spatial profile of the perpendicular magnetization component, $m_z$, as shown in Fig.~\ref{Fig:Trajectory}(d), at the largest value of $j_z < j_z^\mathrm{th}$ considered in simulation. The numerical values of $\kappa^\mathrm{c}$ we obtained are shown in the inset of of Fig. \ref{Fig:Fbc}(b) as a function of $\alpha$. As expected, the critical size does not depend on $\alpha$ and we find $\kappa^\mathrm{c} = \left( 37.7 \pm 0.3 \right)$ nm in this model. Along with the numerical estimates of $j_z^\mathrm{th}$, we can determine $F_\mathrm{b}^\mathrm{c}$ from
\begin{equation} \label{Eq:Fbcz}
	F_\mathrm{b}^\mathrm{c} = -\frac{ G \sigma \kappa^\mathrm{c}}{\mathcal{D} \alpha } j_z^\mathrm{th}.
\end{equation}
The result is plotted in Figure \ref{Fig:Fbc}(b) which shows that $\left| F_\mathrm{b}^\mathrm{c} \right|$ for the $j_z$-driven skyrmion expulsion is also not dependent on $\alpha$ and the value is very close to $\left| F_\mathrm{b}^\mathrm{c} \right|$ obtained from the $j_x$-driven skyrmion expulsion.

\begin{figure}
\centering
\includegraphics[width=8.6cm]{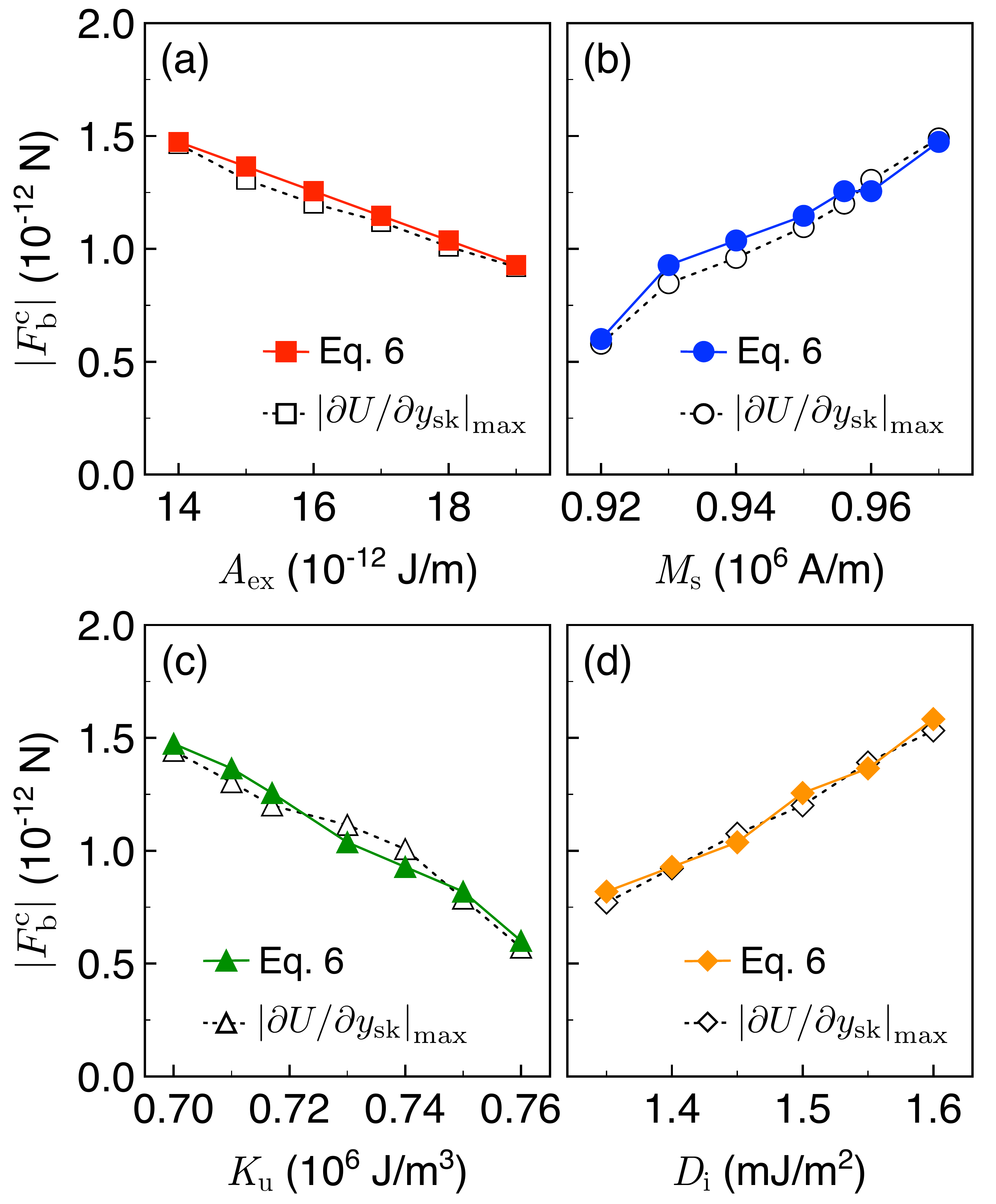}
\caption{\label{Fig:Fbc_var}$\left| F_{\mathrm{b}}^\mathrm{c} \right|$ versus (a) $A_\mathrm{ex}$, (a) $M_\mathrm{s}$, (a) $K_\mathrm{u}$, and (d) $D_\mathrm{i}$ calculated by Eq. \ref{Eq:Fbc} (red squares). The open symbols in (a) - (d) indicate $\left| \partial U / \partial y_\mathrm{sk} \right|_\mathrm{max}$ versus the magnetic parameters. In these calculations, $\alpha$ = 0.3 and  $\beta$ = 0.1 are used.
}
\end{figure}

From the results in Figs. \ref{Fig:Fbc}(a) and \ref{Fig:Fbc}(b), we find that $F_\mathrm{b}^\mathrm{c} = \left( 1.215 \pm 0.036 \right) \times 10^{-12}$ N in our system, which is independent of  $\alpha$ and $\beta$, and almost identical for both $j_x$- and $j_z$-driven skyrmion expulsion. By using this value of the critical boundary force, we are able to determine the $j_x^\mathrm{th}$ and $j_z^\mathrm{th}$ using Eqs. \ref{Eq:jxth} and \ref{Eq:jzth}, respectively, which are in good agreements with the simulation results, as shown in Figs. \ref{Fig:jth}(a) and \ref{Fig:jth}(b). However, this critical boundary force does depend on the magnetic parameters, $A_\mathrm{ex}$, $M_\mathrm{s}$, $K_\mathrm{u}$, and $D_\mathrm{i}$. Figures~\ref{Fig:Fbc_var}(a) - ~\ref{Fig:Fbc_var}(d) show the dependence of $F_\mathrm{b}^\mathrm{c}$ on these magnetic parameters (closed symbols with solid lines). 
In the calculation, $j_x$ is used for driving the skyrmion motion and the range of magnetic parameters are chosen such that the skyrmion state remains stable. We find that $F_\mathrm{b}^\mathrm{c}$ monotonically decreases with increasing $A_\mathrm{ex}$ and $K_\mathrm{u}$, while it increases with increasing $M_\mathrm{s}$ and $D_\mathrm{i}$ in the given parameter ranges.

\subsection{Physical interpretation of the critical boundary force}
\begin{figure}
\centering
\includegraphics[width=8.6cm]{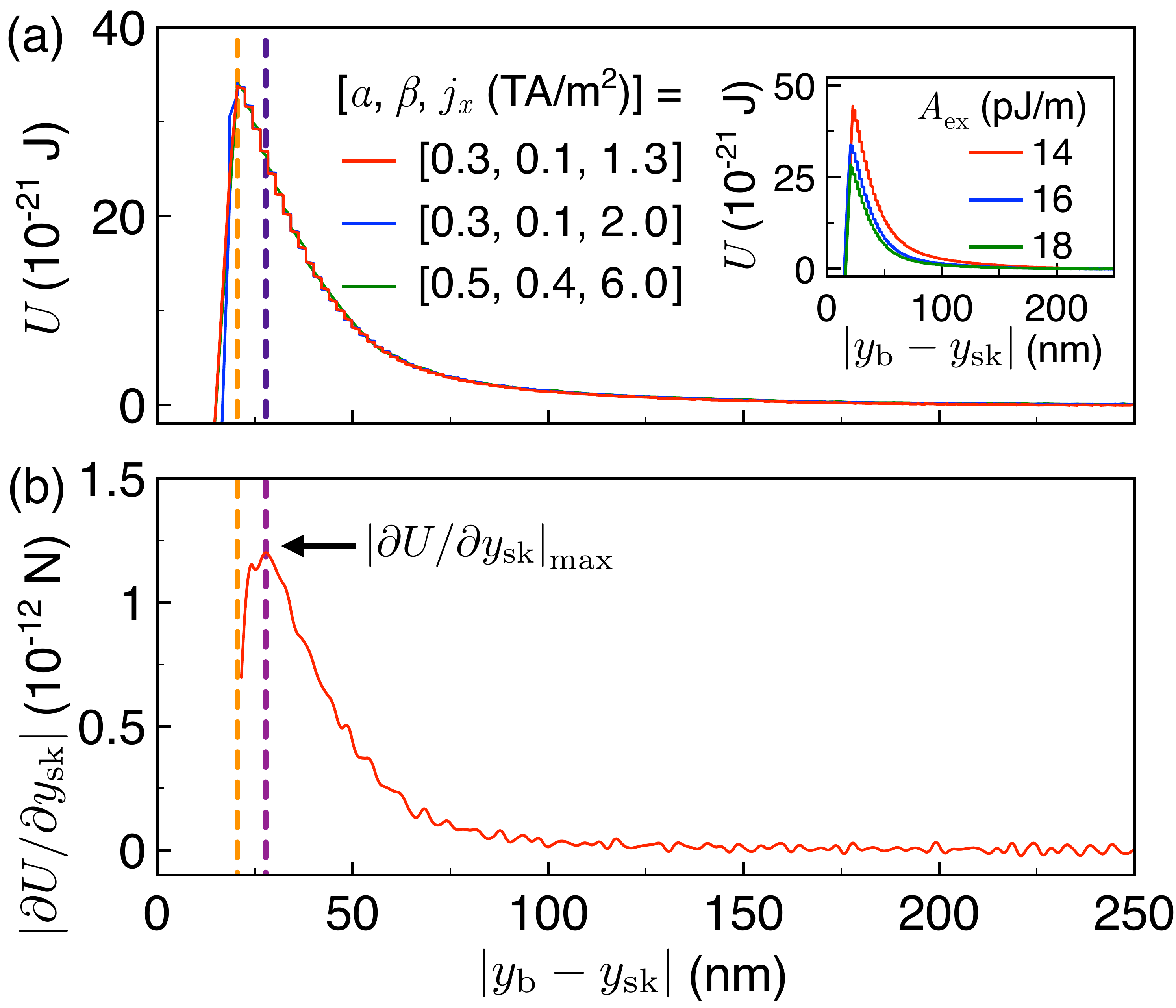}
\caption{\label{Fig:Slope}
(a) $U$ versus $|y_\mathrm{b} - y_\mathrm{sk}|$ obtained from micromagnetic simulations with different values of $\alpha$, $\beta$, and $j_x$, where $y_\mathrm{b}$ is the position of the boundary of the nanowire. The inset shows $U$ versus $|y_\mathrm{b} - y_\mathrm{sk}|$ for different $A_\mathrm{ex}$. (b) $|\partial{U} / \partial{y_\mathrm{sk}}|$ versus $|y_\mathrm{b} - y_\mathrm{sk}|$ numerically obtained from the result of $\alpha$ = 0.3, $\beta$ = 0.1, and $j_x$ = 1.3 $\mathrm{TA/m^2}$ in (a) before the skyrmion expulsion. $|\partial{U} / \partial{y_\mathrm{sk}}|_\mathrm{max}$ indicates the maximum value of $|\partial{U} / \partial{y_\mathrm{sk}}|$. Orange- and violet-colored dashed lines in (a) and (b) indicate the positions of the maximum $U$ and $|\partial{U} / \partial{y_\mathrm{sk}}|_\mathrm{max}$, respectively.
}
\end{figure}

In the Thiele equation, $\mathbf{F}_\mathrm{b}$ is defined as the gradient of $U(\mathbf{X}_\mathrm{sk})$. In our case, $U(\mathbf{X}_\mathrm{sk}) = U(y_\mathrm{sk})$, because we assume an infinitely extended nanostrip in the $x$-direction. Thus, in order to examine the physical meaning of $\mathbf{F}_\mathrm{b}^\mathrm{c}$, we constructed numerically the function $U(y_\mathrm{sk})$ from the micromagnetic simulations, as shown in Fig. \ref{Fig:Slope}(a). In the calculation, $\alpha$ = 0.3, $\beta$ = 0.1, and $j_x = -1.3 \times 10^{12} \mathrm{A/m^2}$ were chosen such that the skyrmion is eventually expelled at the boundary edge, but the function $U(y_\mathrm{sk})$ does not depend on the parameters, $\alpha$, $\beta$, and $j_x$ [see the blue and green lines in Fig.~\ref{Fig:Slope}(a)].
Figure~\ref{Fig:Slope}(a) shows that $U(y_\mathrm{sk})$ has an almost constant value when the skyrmion is far enough from the boundaries, $y_\mathrm{b}$. However, when the skyrmion is sufficiently close to the boundary edge and $\left| y_\mathrm{b} - y_\mathrm{sk} \right|$ becomes smaller, $U(y_\mathrm{sk})$ increases sharply as a result of the interaction between the skyrmion and the boundary.
When $\left| y_\mathrm{b} - y_\mathrm{sk} \right|$ reaches a certain critical value [orange dashed line in Fig. \ref{Fig:Slope}(a)], $U(y_\mathrm{sk})$ attains a maximum and then decreases drastically as the skyrmion is expelled from the nanowire. From the function of $U(y_\mathrm{sk})$, the gradient, $\partial{U} / \partial{y_\mathrm{sk}}$, can be obtained numerically [Fig \ref{Fig:Slope}(b)]. $\left| \partial{U} / \partial{y_\mathrm{sk}} \right|$ has the maximum value, $\left| \partial{U} / \partial{y_\mathrm{sk}} \right|_\mathrm{max}$, just before the expulsion [violet dashed line in Fig. \ref{Fig:Slope}(b)], and the value of $|\partial{U} / \partial{y_\mathrm{sk}}|_\mathrm{max} =  1.200 \times 10^{-12}$ N is very close to $\left| F_\mathrm{b}^\mathrm{c} \right| = \left(1.215 \pm 0.036 \right) \times 10^{-12}$ N obtained from the simulations with Eqs.~(\ref{Eq:jxth}) and (\ref{Eq:jzth}).

The function $U(\mathbf{X}_\mathrm{sk})$ is strongly dependent on the magnetic parameters, as shown in the inset of Fig. \ref{Fig:Slope}(a). We also calculated $|\partial{U} / \partial{y_\mathrm{sk}}|_\mathrm{max}$ for different magnetic parameters and plotted them in Figs. \ref{Fig:Fbc_var}(a) - \ref{Fig:Fbc_var}(d) (open symbols with dashed lines), which are in good agreements qualitatively and quantitatively with $\left| F_\mathrm{b}^\mathrm{c} \right|$s obtained from Eqs. \ref{Eq:jxth} and \ref{Eq:jzth}. These results clearly show that the maximum gradient of $U(y_\mathrm{sk})$ before the expulsion, $\left |\partial{U} / \partial{y_\mathrm{sk}} \right|_\mathrm{max}$, corresponds to $\left| F_\mathrm{b}^\mathrm{c} \right|$, as expected from the Thiele equation, and the magnetic-parameter dependence of $\left| F_\mathrm{b}^\mathrm{c} \right|$ originates from the dependence of $U(\mathbf{X}_\mathrm{sk})$ on the magnetic parameters.

\subsection{Model of the skyrmion-boundary interaction}
The potential $U\left(y_\mathrm{sk}\right)$ results from the interaction between the skyrmion and the boundary of the nanostrip. Here, we present a simple model to describe this interaction by considering how a skyrmion is repelled by a partial N{\'e}el domain wall, which describes the magnetization tilt at the boundary edge. This tilt can be seen in Fig. \ref{Fig:InitialState} (orange-colored region), where the magnetization near the edge deviates from the easy axis ($m_z = -1$) direction as a result of the DMI-induced boundary condition, $D_\mathrm{i} m_z + 2A_\mathrm{ex} (\partial{m_y}/\partial{y}) = 0$ and $-D_\mathrm{i} m_y + 2A_\mathrm{ex} (\partial{m_z}/\partial{y}) = 0$ \cite{Rohart:2013ef, GarciaSanchez:2014dw}. The nonuniform magnetization near the boundary can be described by a partially expelled N{\'e}el-type domain wall~\cite{GarciaSanchez:2014dw}, $\mathbf{m}_\mathrm{dw} = \left( m_{x,dw}, m_{y,dw}, m_{z,dw} \right) = \left( \cos{\phi_\mathrm{dw}} \sin{\theta_\mathrm{dw}}, \sin{\phi_\mathrm{dw}} \sin{\theta_\mathrm{dw}}, \cos{\theta_\mathrm{dw}} \right)$, where
\begin{subequations} \label{Eq:mdw}
  \begin{align}
    \theta_\mathrm{dw} &= \pm \arccos \left[ \tanh{\left(\tilde{y}-\tilde{y}_\mathrm{c}\right)} \right] \label{thdw} \\
    \mathrm{and} \quad \phi_\mathrm{dw} &= \pm \frac{\pi}{2} \label{phidw}
  \end{align}
\end{subequations}
are the polar and azimuthal angles of the local magnetization vector, respectively. In this calculation, we use the characteristic length scale $\lambda = \sqrt{A_\mathrm{ex}/K_\mathrm{0}}$ in order to define the dimensionless spatial variables, $\tilde{x} = x/\lambda$, $\tilde{y} = y/\lambda$, and $\tilde{z} = z/\lambda$, respectively. In Eq.~(\ref{Eq:mdw}), $\tilde{y}_\mathrm{c} = \tilde{y}_\mathrm{b} \pm \mathrm{arcsech} \left( D_\mathrm{0}/2 \right)$ is the center of the domain wall which is located outside of the nanostrip, where $D_0 = D_\mathrm{i}/\sqrt{A_\mathrm{ex} K_0}$ and $K_0 = K_\mathrm{u} - \mu_0 M_\mathrm{s}^2/2$.

The configuration of an isolated skyrmion can be described using the double-soliton ansatz,  $\mathbf{m}_\mathrm{sk} = \left( m_{x,sk}, m_{y,sk}, m_{z,sk} \right) = \left( \cos{\phi_\mathrm{sk}} \sin{\theta_\mathrm{sk}}, \sin{\phi_\mathrm{sk}} \sin{\theta_\mathrm{sk}}, \cos{\theta_\mathrm{sk}} \right)$~\cite{Braun:1994ff, Romming:2015il,GarciaSanchez:2016cxa}, where 
\begin{subequations} \label{Eq:msk}
  \begin{align}
	\theta_\mathrm{sk} &= \pm \arccos {\left( \frac{4 \cosh^2{\tilde{c}}}{\cosh{(2\tilde{c})}+\cosh{(2\tilde{r})}}-1 \right)} \label{thetask} \\
	\mathrm{and} \quad \phi_\mathrm{sk} &= \frac{\pi}{2} + \arctan {\left( \frac{\tilde{y}-\tilde{y}_\mathrm{sk}}{\tilde{x}-\tilde{x}_\mathrm{sk}} \right)} \pm \frac{\pi}{2}.
  \end{align}
\end{subequations}
In Eq. \ref{Eq:msk}, $\tilde{r} = \sqrt{(\tilde{x}-\tilde{x}_\mathrm{sk})^2+(\tilde{y}-\tilde{y}_\mathrm{sk})^2}$ is a distance from the skyrmion center and $\tilde{c}$ is a distance between two successive $180^{\circ}$ homochiral domain-walls that is proportional to the size of the skyrmion. The ($\pm$) signs in Eqs. \ref{Eq:mdw} and \ref{Eq:msk} are determined by the saturation direction of the given nanowire and the sign of $D_\mathrm{i}$. Note that we have assumed a fixed domain-wall width $\lambda = \sqrt{A_\mathrm{ex}/K_\mathrm{0}}$, in both $\mathbf{m}_\mathrm{dw}$ and $\mathbf{m}_\mathrm{sk}$.\par

\begin{figure}
\centering\includegraphics[width=8.6cm]{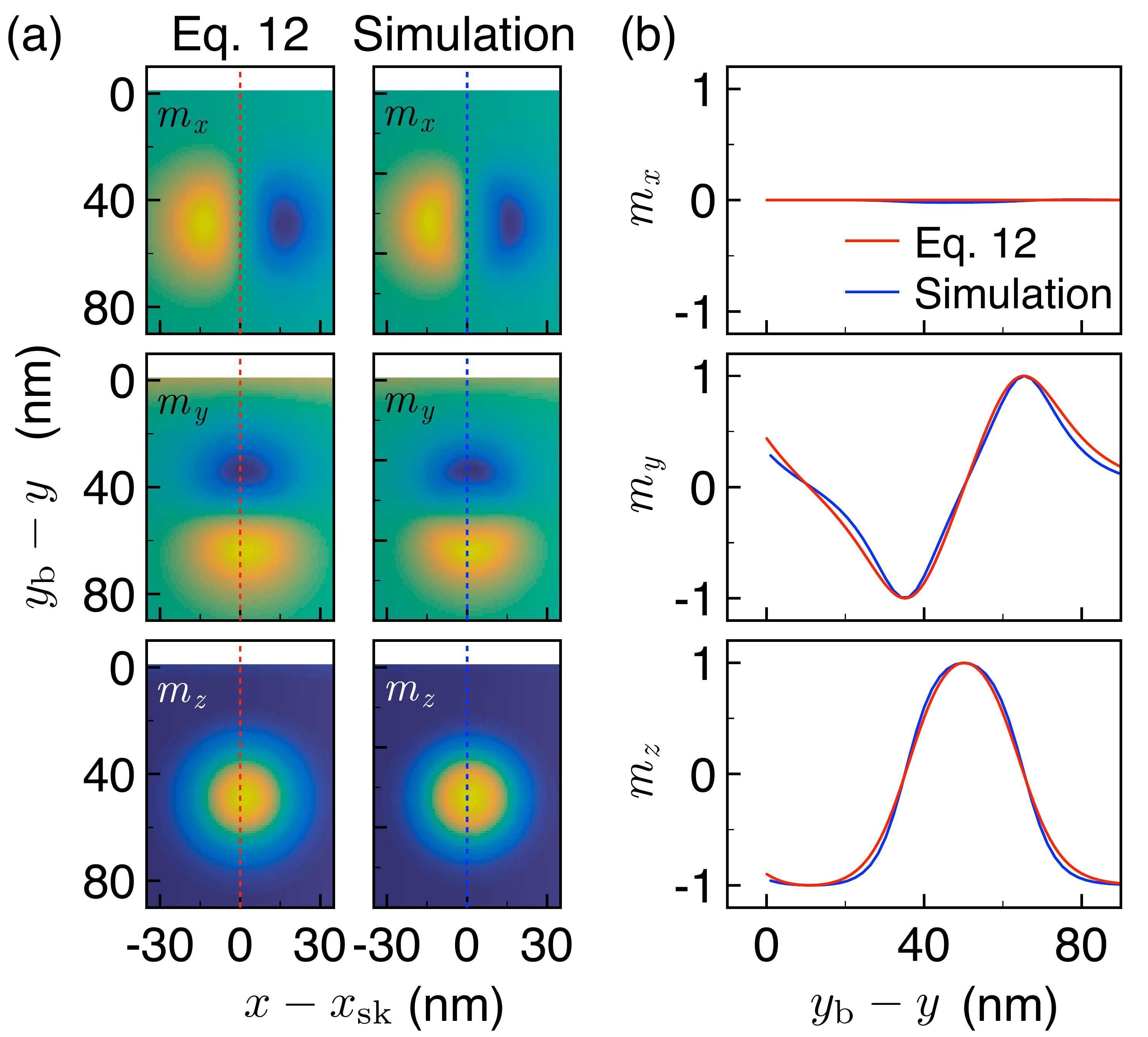}
\caption{\label{Fig:Analytic_Model}(a) $m_x$, $m_y$, and $m_z$ distributions obtained from Eq. \ref{Eq:mm} (left panels) and micromagnetic simulations (right panels). For the analytic model (left panels), we used $\tilde{y}_\mathrm{sk}$ = 4.70 and $\tilde{c}$ = 1.30 which values correspond to the simulation result (right panels). (b) $m_x$, $m_y$, and $m_z$ profiles of the skyrmion and the edge of nanowire along the $y$-direction (dashed lines in (a)).
}
\end{figure}

To describe the interaction between the skyrmion and the partial N{\'e}el wall, which represents the boundary edge, we construct a superposition of the two spin textures $\mathbf{m}_{\mathrm{sk}}$ and $\mathbf{m}_{\mathrm{dw}}$ in the following way,
\begin{equation} \label{Eq:mm}
  \mathbf{m}=\frac{w_\mathrm{sk} \mathbf{m}_\mathrm{sk} + w_\mathrm{dw} \mathbf{m}_\mathrm{dw}}{\left| w_\mathrm{sk} \mathbf{m}_\mathrm{sk} + w_\mathrm{dw} \mathbf{m}_\mathrm{dw} \right|}.
\end{equation}
$w_\mathrm{sk} = | \theta_\mathrm{s} - \theta_\mathrm{sk} |/w$ and $w_\mathrm{dw} = | \theta_\mathrm{s} - \theta_\mathrm{dw} |/w$  are weighting functions of each spin texture, where $w = | \theta_\mathrm{s} - \theta_\mathrm{sk} | + | \theta_\mathrm{s} - \theta_\mathrm{dw} |$. The values of $w_\mathrm{sk}$ and $w_\mathrm{dw}$ are proportional to the deviation of the local magnetization from the saturation orientation of the magnetization in the nanowire, $\theta_{\mathrm{s}} = \pi$ or $0$. These weights are necessary since a simple superposition of the spin textures, $\mathbf{m}_{\mathrm{sk}}+ \mathbf{m}_{\mathrm{dw}}$, would not preserve the condition on the norm of the magnetization field, $\| \mathbf{m} \|  =1$. In Fig.~\ref{Fig:Analytic_Model}, we compare this analytical model with results from micromagnetic simulations.
First, we obtained the magnetic configuration from the simulation at certain skyrmion position ($\tilde{y}_\mathrm{sk}$ = 4.70) and skyrmion size ($\tilde{c}$ = 1.30). Using the obtained values of $\tilde{y}_\mathrm{sk}$ and $\tilde{c}$, Eq. \ref{Eq:mm} is calculated, and $m_x$, $m_y$, and $m_z$ are displayed in Fig. \ref{Fig:Analytic_Model} as well as those obtained from the micromagnetic simulations. As shown in Fig.~\ref{Fig:Analytic_Model}, we found that the skyrmion-boundary model, Eq.~\ref{Eq:mm}, provides a good description of the magnetic configuration of the skyrmion near the edge as well as that of the boundary.

Based on this model, we can compute the potential energy $U$ as a function of $y_\mathrm{sk}$. In order to simplify the calculation, we assume a local form for the dipolar interaction and use an energy scale of $U_0 = A_{\mathrm{ex}}d$,  $U/U_0 = u$. By assuming $\tilde{y}_\mathrm{b} > \tilde{y}_\mathrm{sk}$, the total magnetic energy, $u$, can be calculated by $u = u_\mathrm{ex} + u_\mathrm{ani} + u_\mathrm{dmi}$, where
\begin{subequations} \label{Eq:uu}
  \begin{align}
    u_\mathrm{ex} &= \int_{-\infty}^\infty \int_{-\infty}^{\tilde{y}_\mathrm{b}} \left( \nabla \mathbf{m} \right)^2 d \tilde{y} d \tilde{x},\\
    u_\mathrm{ani} &= \int_{-\infty}^\infty \int_{-\infty}^{\tilde{y}_\mathrm{b}} {m_z}^2 d \tilde{y} d \tilde{x}, \qquad{} \mathrm{and}\\ 
    \begin{split}
      u_\mathrm{dmi} &= D_0 \int_{-\infty}^\infty \int_{-\infty}^{\tilde{y}_\mathrm{b}} \left( m_z \frac{\partial m_x}{\partial \tilde{x}}-m_x \frac{\partial m_z}{\partial \tilde{x}} \right)\\ &\qquad{}\qquad{}\qquad{}+ \left( m_z \frac{\partial m_y}{\partial \tilde{y}}-m_y \frac{\partial m_z}{\partial \tilde{y}} \right) d \tilde{y} d \tilde{x}
    \end{split}
  \end{align}
\end{subequations}
are the Heisenberg exchange, anisotropy, and DMI energies, respectively. Note that $u$ in Eq. \ref{Eq:uu} is only a function of  $\tilde{c}$, $\tilde{y}_\mathrm{b}-\tilde{y}_\mathrm{sk}$, and $D_{\mathrm{0}}$, i.e., $u\left(\tilde{c}, \tilde{y}_\mathrm{b}-\tilde{y}_\mathrm{sk} ,D_{\mathrm{0}}\right)$, and, in our case, $D_0 \sim 0.9925$ for the chosen magnetic parameters.

\begin{figure}
\centering\includegraphics[width=8.6cm]{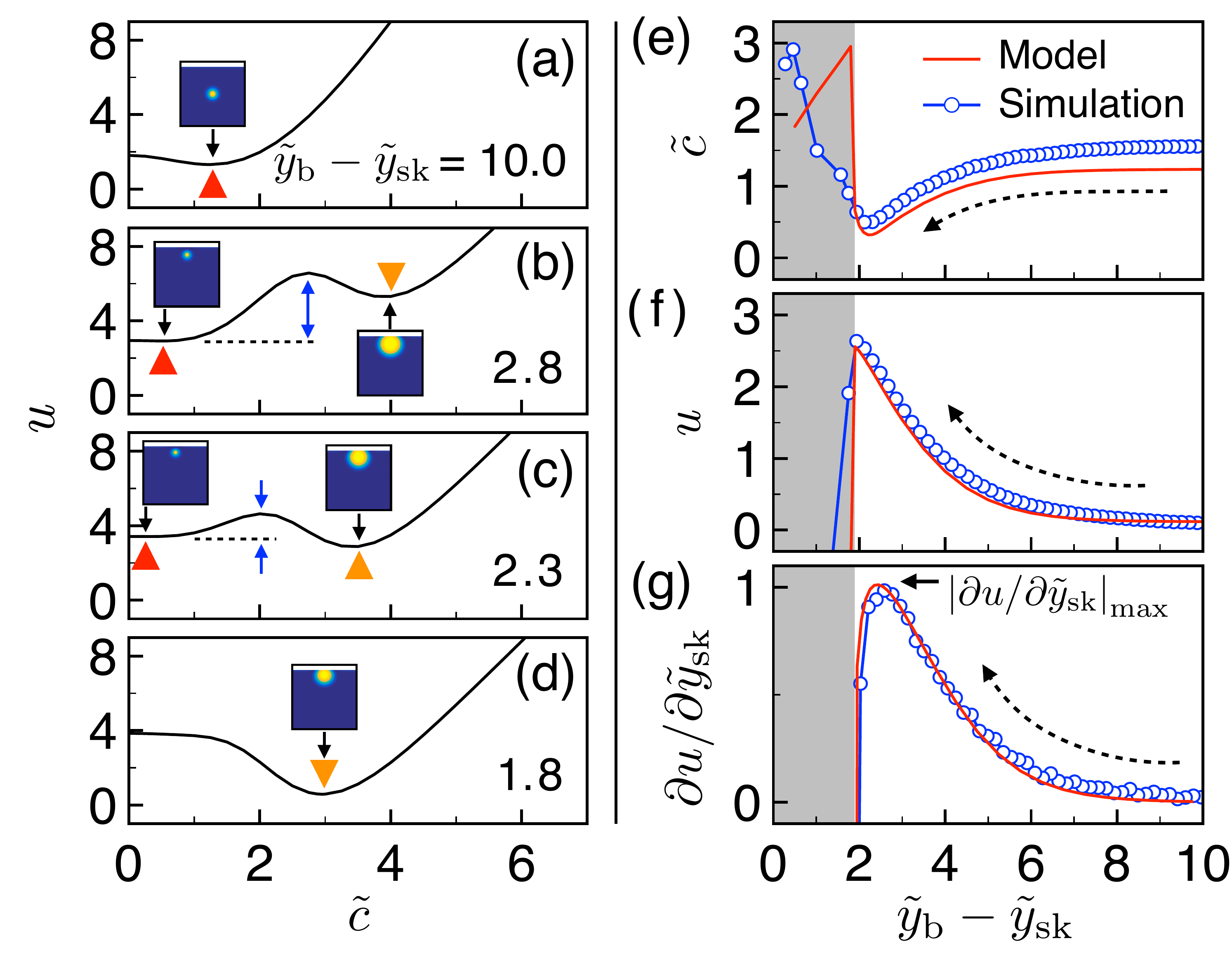}
\caption{\label{Fig:uu}
(a) - (d) $u$ versus $\tilde{c}$ calculated from Eq. \ref{Eq:uu} for different values of $\tilde{y}_\mathrm{b}-\tilde{y}_\mathrm{sk}$. The red and orange triangles indicate the local minimum states of the skyrmion and partially expelled skyrmion, respectively. The blue arrows in (c) and (d) represents the energy barriers between the two minimum states. The insets show the magnetic configurations at the minimum states. (e) Most stable $\tilde{c}$, (f) $u$, and (g) $\partial{u}/\partial{\tilde{y}_\mathrm{sk}}$ versus $\tilde{y}_\mathrm{b}-\tilde{y}_\mathrm{sk}$ obtained from the skyrmion-boundary model (red lines) and those obtained from the micromagnetic simulations (blue open circles). The gray-colored regions correspond to the partially expelled skyrmion state obtained from the skymion-boundary model. The dashed arrows represent the approaching direction of the skyrmion to the boundary.
}
\end{figure}

By using Eq.~\ref{Eq:uu} and the given $D_0$, $u$ as a function of $\tilde{c}$ can be calculated at finite values of $\tilde{y}_\mathrm{b}-\tilde{y}_\mathrm{sk}$, and, from the $u$-$\tilde{c}$ relations, the most stable $\tilde{c}$ can be obtained by $\partial{u}/\partial{\tilde{c}} = 0$ [Fig.~\ref{Fig:uu}(a)]. 
When the skyrmion is sufficiently far from the boundary, the system only has one minimum energy state [Fig.~\ref{Fig:uu}(a)], which corresponds to the stable isolated skyrmion state in an infinite magnetic film; the value of $\tilde{c}$ almost does not vary with $\tilde{y}_\mathrm{b}-\tilde{y}_\mathrm{sk}$, when $\tilde{y}_\mathrm{b}-\tilde{y}_\mathrm{sk} \gg 0$. As the skyrmion approaches the edge, the stable $\tilde{c}$ configuration gradually decreases and another minimum state, the partially expelled skyrmion state, appears at a larger value of $\tilde{c}$ [orange triangle in Fig. \ref{Fig:uu}(b)]. The energy of the partially expelled skyrmion states decreases with decreasing $\tilde{y}_\mathrm{b}-\tilde{y}_\mathrm{sk}$, and it becomes more stable than the whole skyrmion state [Fig. \ref{Fig:uu}(c)]. In this calculation, however, a zero temperature is assumed, thus the skyrmion state [red triangle in Fig. \ref{Fig:uu}(c)] cannot overcome the energy barrier to another minimum energy states. Finally, when the energy barrier between the two minimum states disappears, the skyrmion is expelled from the magnetic nanowire [Fig.~\ref{Fig:uu}(d)]. The most stable $\tilde{c}$ value and the corresponding $u$ are plotted in Figs. \ref{Fig:uu}(e) and \ref{Fig:uu}(f), respectively, as a function of $\tilde{y}_\mathrm{b}-\tilde{y}_\mathrm{sk}$. We find good agreement with the variation in $\tilde{c}$ and $u$ obtained from micromagnetics simulations before the skyrmion is expelled. We note that, for the skyrmion-boundary model, we consider the dipolar energy as a local approximation that affects the calculated skyrmion energy and skyrmion size. If we consider the dipolar coupling without this approximation, the results would be more agreement with the simulations, however, the difference is negligibly small, as shown in Fig.~\ref{Fig:Analytic_Model} and Fig.~\ref{Fig:uu}, because the approximation is quite valid for the ultrathin system~\cite{Rohart:2013ef}.
\begin{figure}
\centering
\includegraphics[width=8.6cm]{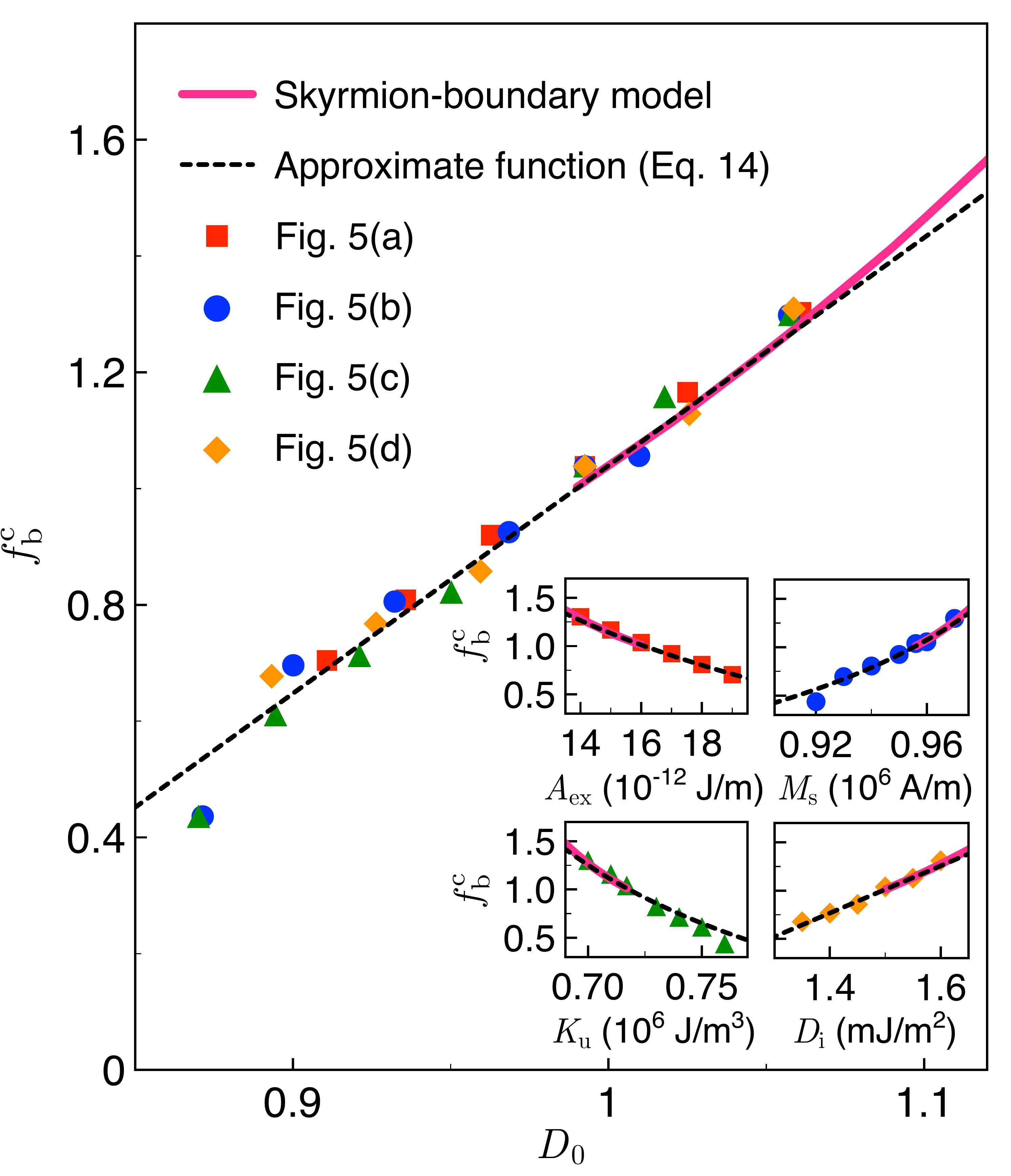}
\caption{\label{Fig:FbcD0}
$f_\mathrm{b}^\mathrm{c}$ obtained from the skyrmion-boundary model (solid line) and approximate linear function, $f_{\mathrm{b}}^{\mathrm{c}} = \eta \left( D_0 - 1 \right) + f_\mathrm{b}^\mathrm{c} \left( 1 \right)$ (Eq. \ref{Eq:fbc}), where $\eta$ = 3.699 and $f_\mathrm{b}^\mathrm{c} \left( 1 \right)$ = 1.04 (dashed line). The symbols are $f_\mathrm{b}^\mathrm{c}$s obtained from the micromagnetic simulations, which correspond to the data in Figs.~\ref{Fig:Fbc_var}(a) - \ref{Fig:Fbc_var}(d) represented by the filled symbols.  The inset shows $f_{\mathrm{b}}^\mathrm{c}$ versus $A_\mathrm{ex}$, $M_\mathrm{s}$, $K_\mathrm{u}$, and $D_\mathrm{i}$.
}
\end{figure}

From the relationship between $u$ and $\left(\tilde{y}_\mathrm{b}-\tilde{y}_\mathrm{sk}\right)$, the critical boundary force, $f_\mathrm{b}^\mathrm{c} = F_\mathrm{b}^\mathrm{c} \lambda / U_0$, can be calculated from the maximum gradient value, $\left| \partial{u}/\partial{\tilde{y}_{sk}} \right|_\mathrm{max}$, before the partial expulsion. As shown in Fig. \ref{Fig:uu}(g), the obtained value $f_\mathrm{b}^\mathrm{c}$ = $\left| \partial{u}/\partial{y_{sk}} \right|_\mathrm{max}$ = 1.012 is in a good agreement with the simulation result $f_\mathrm{b}^\mathrm{c}$ = 1.005 presented in Fig. \ref{Fig:Fbc}. Finally, $f_\mathrm{b}^\mathrm{c}$ for different $D_0$ are calculated and plotted in Fig. \ref{Fig:FbcD0}. Note that $f_\mathrm{b}^\mathrm{c}$ cannot be obtained accurately from the skyrmion-boundary model for $D_0 \leq 0.98$, since for these values the skyrmion size before expulsion is below that given by $\tilde{c} = 0$. Figure~\ref{Fig:FbcD0}(a) shows that $f_\mathrm{b}^\mathrm{c}$ increases monotonically with increasing $D_0$. Near $D_0 \sim 1$, the function $f_\mathrm{b}^\mathrm{c} \left( D_0 \right)$ shows a quasi-linear behavior which can be approximately expressed by a linear function as 
\begin{equation} \label{Eq:fbc}
    f_{\mathrm{b}}^{\mathrm{c}} \left( D_0 \right) = \eta \left( D_0 - 1 \right) + f_{\mathrm{b}}^{\mathrm{c}} \left( 1 \right),
\end{equation}
where $\eta$ = $\partial{f_\mathrm{b}^\mathrm{c}}/\partial{D_0}|_{D_0 = 1} \sim 3.699$ and $f_\mathrm{b}^\mathrm{c} \left( 1 \right)$ = 1.04. As shown in Fig. \ref{Fig:FbcD0}, the function of $f_\mathrm{b}^\mathrm{c}$ and the approximate function (Eq. \ref{Eq:fbc}) are in good agreements with the simulation results in Fig. \ref{Fig:Fbc_var} in the range of $0.9 < D_0 < 1.1$, and clearly explains the dependences of $f_\mathrm{b}^\mathrm{c}$ on the magnetic parameters: $A_\mathrm{ex}$, $M_\mathrm{s}$, $K_\mathrm{u}$, and $D_\mathrm{i}$ [The inset of Fig. \ref{Fig:FbcD0}]. From this result, we find that $D_0$ is the key parameter for determining $f_\mathrm{b}^\mathrm{c}$, and $D_0$ is the origin of the dependence of $F_\mathrm{b}^\mathrm{c}$ on the magnetic parameters presented in Fig. \ref{Fig:FbcD0}.

\section{Conclusion}
We have presented a theoretical study of current-driven skyrmion expulsion in magnetic nanostrips. A finite current threshold exists for this expulsion because magnetization tilts at the boundary edge result in a confining potential that acts to keep the skyrmion within the nanostrip. The threshold current density for the expulsion depends on the critical boundary force as well as the spin torque parameters, such as the Gilbert damping constant and/or the non-adiabaticity parameter. The critical boundary force is found to depend on the scaled DMI parameter, $D_0 = D_\mathrm{i}/\sqrt{A_\mathrm{ex} K_0}$. A linear approximation for the critical boundary force as a function of $D_0$ is found to describe well the simulation results for a range of values around $D_0 \sim 1$. This work provides a fundamental understanding of the skyrmion-boundary interaction as well as skyrmion expulsion, and shows that the stability of the skyrmion at the boundaries of devices can be related to $D_0$ of magnetic materials.

\begin{acknowledgments}
The authors would like to acknowledge fruitful discussions with Stanislas Rohart. We also would like to acknowledge a careful reading and valuable comments by Nicolas Reyren. This work was supported by the Horizon2020 Framework Programme of the European Commission, under grant agreement No. 665095 (MAGicSky).
\end{acknowledgments}

\bibliography{skyrmion_expulsion}{}
\bibliographystyle{apsrev4-1}

\end{document}